\documentclass[twocolumn,superscriptaddress,longbibliography]{revtex4-1}%
\usepackage{graphicx}
\usepackage{dcolumn}
\usepackage{bm}
\usepackage[dvipsnames]{xcolor} 
\usepackage{hyperref} 
\usepackage{amsmath}%
\usepackage{amsfonts}%
\usepackage{amssymb}

\hypersetup{
    colorlinks=true,
    linkcolor=Blue,
    filecolor=magenta,
    urlcolor=Black,
    citecolor=Black,
}
\usepackage{float}

\begin{document}
\title{\textbf{Microscopic picture of electron-phonon interaction in two-dimensional halide perovskites}}
\author{David Feldstein}
\author{Ra\"ul Perea-Caus\'in}

\affiliation{Chalmers University of Technology, Department of Physics,
 412 96 Gothenburg, Sweden}

 \author{Shuli Wang}
 \affiliation{Laboratoire National des Champs Magn\'etiques Intenses, UPR 3228, CNRS-UGA-UPS-INSA, Grenoble and Toulouse, France}
 \author{Mateusz Dyksik}
\affiliation{Laboratoire National des Champs Magn\'etiques Intenses, UPR 3228, CNRS-UGA-UPS-INSA, Grenoble and Toulouse, France}
\affiliation{Department of Experimental Physics, Faculty of Fundamental Problems of Technology, Wroclaw University of Science and Technology, Wroclaw, Poland}

\author{Kenji Watanabe}
\author{Takashi Taniguchi}

\affiliation{Research Center for Functional Materials,
National Institute for Materials Science, 1-1 Namiki, Tsukuba 305-0044, Japan}

\author{Paulina Plochocka}
 \affiliation{Laboratoire National des Champs Magn\'etiques Intenses, UPR 3228, CNRS-UGA-UPS-INSA, Grenoble and Toulouse, France}
\affiliation{Department of Experimental Physics, Faculty of Fundamental Problems of Technology, Wroclaw University of Science and Technology, Wroclaw, Poland}

\author{Ermin Malic}

\affiliation{Chalmers University of Technology, Department of Physics,
 412 96 Gothenburg, Sweden}

\begin{abstract}
Perovskites have attracted much attention due to their remarkable optical properties. While it is well established that excitons dominate their optical response, the impact of higher excitonic states and formation of phonon sidebands in optical spectra still need to be better understood. Here, we perform a theoretical study on excitonic properties of monolayered hybrid organic perovskites---supported by temperature-dependent photoluminescence measurements. Solving the Wannier equation, we obtain microscopic access to the Rydberg-like series of excitonic states including their wavefunctions and binding energies.
Exploiting the generalized Elliot formula, we calculate the photoluminescence spectra demonstrating a pronounced contribution of a phonon sideband for temperatures up to  50 K---in agreement with experimental measurements. Finally, we predict temperature-dependent linewidths of the three energetically lowest excitonic transitions and identify the underlying phonon-driven scattering processes.\\
\end{abstract}

\maketitle


In the last years, much effort has been devoted to the study of perovskite materials\,\cite{jena19,stoumpos13,niu15,yin15,Tsai2016}. Their broad absorption spectrum\,\cite{stoumpos14} combined with impressive light-emitting properties\,\cite{sampson} make them a promising material for optoelectronic applications. They have been, in particular, applied as active material in solar devices exhibiting a power conversion efficiency of up to  25.2 \%\,\cite{nrel,huang15}. Beyond photovoltaics there has been a rising interest in perovskite based diodes and lasers\,\cite{Tan2014,Cho1222,Matsushima2019, Gong2018}.
Recently, layered perovskites have emerged as a subclass of the bulk material exhibiting enhanced excitonic properties and tunable characteristics\,\cite{stoumpos19}. Their optical and electronic properties can be tailored through an external stress\,\cite{Mante2017,lu17}, an internal chemical pressure\,\cite{sun15} or by varying their layer thickness\,\cite{etgar18}. In addition, a wide variety of compositions can be made thanks to the large amount of organic cations that can be integrated into the structure. These materials are considered as a dimensional reduction of 3D perovskites with the general formula (A)$'_m$A$_{n-1}$B$_n$X$_{3n+1}$, where A$'$ is the cation that intercalates between the inorganic A$_{n-1}$B$_n$X$_{3n+1}$ layers and $n$ can be understood as the thickness of the inorganic layers. Here, the screening of the Coulomb interaction is weaker compared to 3D perovskites resulting in considerably larger exciton binding energies\,\cite{milot16,yaffe15}. As a direct consequence, excitons are stable at room temperature and determine the optical response and the non-equilibrium dynamics of these materials.
Recent experimental studies on optical properties of 2D perovskites have focused on exciton binding energies\,\cite{yaffe2015excitons}, absorption linewidths\,\cite{straus2019longer,ziegler2020fast,guo2016electron,ni2017real,gong2018electron,gauthron2010optical}, phonon sidebands\,\cite{straus}, and high-density effects\,\cite{palmieri2020mahan}. However, a microscopic understanding of some of the reported findings, e.g. the formation of phonon sidebands and the microscopic origin of excitonic linewidths, have not been well understood yet.

\begin{figure}[t!]
 \includegraphics[width=0.85\linewidth]{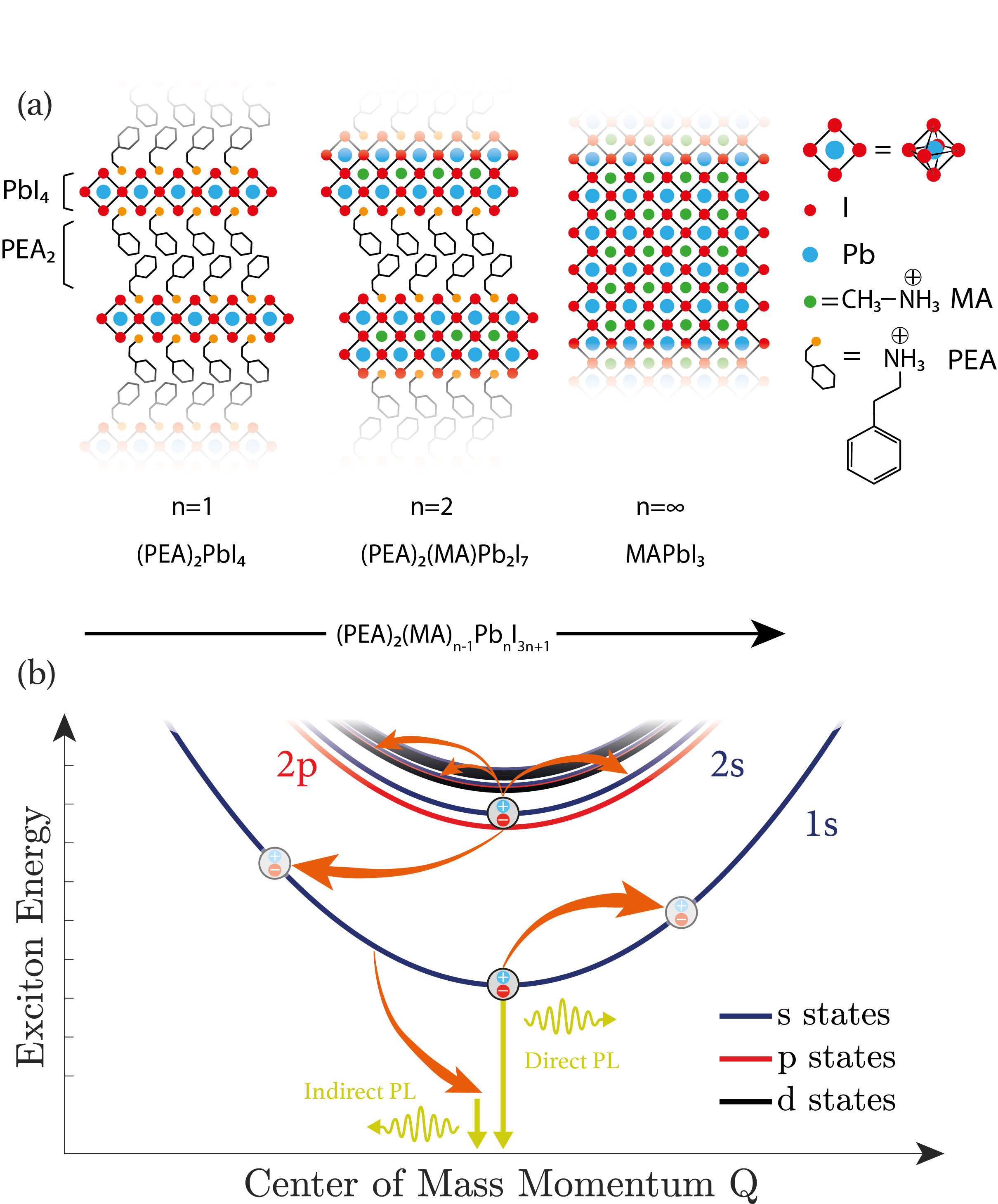}
 \caption{(a) Schematic illustration of the crystal structure of $\textrm{(PEA)}_2 (\textrm{MA})_{n-1} \textrm{Pb}_n \textrm{I}_{3n+1}$  from monolayered (n=1) to bulk perovskites (n=$\infty$).  (b) Exciton bandstructure and possible phonon-assisted scattering channels for 1s and 2s excitons.}
\label{fig:PEA n}
\end{figure}

Here, we present a joint theoretical study on optical properties of monolayered hybrid organic perovskites supported by experimental photoluminescence (PL) measurements. The aim of the work is to provide a microscopic understanding of many-particle processes determining the temperature-dependent PL in these materials. We focus on the monolayered $(\text{PEA)}_2\text{PbI}_4$, which is a particular case of the n-layered $(\text{PEA})_m \text{MA}_{n-1} \text{Pb}_n \text{I}_{3n+1}$ perovskite\,\cite{calabrese,smith14}, cf. Fig.~\ref{fig:PEA n}(a). Here, the inorganic layer consists of corner-sharing $\text{PbI}_6$ octahedra  sandwiched between two organic layers of PEA (phenyl ethyl ammonium). The lower dielectric constant of the organic layer results in a dielectrically confined quantum well.

Using a fully quantum mechanical approach, we obtain microscopic access to excitonic wavefunctions and binding energies in 2D perovskites. Exploiting the generalized Elliot formula, we model temperature-dependent PL spectra including the emergence of phonon-assisted sidebands.  Finally, we determine the spectral linewidth of the energetically lowest excitonic transitions  taking into account scattering channels driven by the emission and absorption of acoustic and optical phonons, cf.   Fig.~\ref{fig:PEA n}(b). The theoretical predictions are compared to temperature-dependent measurements of photoluminescence spectra in the investigated 2D perovskites.\\

As a first step, we determine the excitonic properties of monolayered $(\text{PEA)}_2\text{PbI}_4$ perovskites. To obtain microscopic access to the Rydberg-like series of excitonic states $\nu$, their binding energies $E^{\nu}_b$ and wavefunctions $\phi^{\nu}_{\mathbf{k}}$, we solve the Wannier equation\,\cite{Koch2006, brem2018, selig2018}
\begin{equation}
\label{eqn:excitonic_properties_1}
\frac{\hbar^{2} \mathbf{k}^{2}}{2 \mu}  \phi^{\nu}_{\mathbf{k}} - \sum_{\mathbf{k'}} V_{\mid \mathbf{k}-\mathbf{k'} \mid} \phi^{\nu}_{\mathbf{k'}} = E^{\nu}_b \phi^{\nu}_{\mathbf{k}},
\end{equation}
 where we approximate the Coulomb interaction by the Rytova-Keldysh potential\,\cite{rytova1967,keldysh}
 $
V_{\mathbf{q}}= \frac{e_{0}^2}{2 \varepsilon_0 \varepsilon_s L^2 \mid \mathbf{q} \mid (1+r_{0} \mid \mathbf{q} \mid)}
$
similar to the treatment of atomically thin transition metal dichalcogenides (TMDs)\,\cite{malic_keldysh}.
Here, we used the exciton reduced mass $\mu=0.108m_0$\,\cite{ziegler2020fast} and the average dielectric constant $\varepsilon_s=3.32$ given by the organic layers \,\cite{ishihara}. The dielectric constant of the inorganic layer is $\varepsilon_{\text{well}}=6.1$  and its thickness  $\ L_{\text{well}} = 0.636$ nm\,\cite{ishihara}. These quantities determine the screening length $r_0=L_{\text{well}}\frac{\varepsilon_{\text{well}}}{2\varepsilon_s}$, which is crucial for the excitonic binding energy $E_b$.

\begin{table}[b!]
\centering
\begin{tabular}{|c || c |c | c | c | c | c | c|}
 \hline
 Excitonic state & $E_{1s}$ & $E_{2s}$ & $E_{3s}$ & $E_{4s}$ & $E_{2p}$ & $E_{3p}$ & $E_{3d}$ \\ [1.ex]
 \hline
 Binding energy $E_b$ [meV] & 228 & 43 & 17 & 9 & 53 & 19.5 & 20 \\ [1.ex]
 \hline
\end{tabular}
\caption{Excitonic binding energies $E_b$ obtained by solving the Wannier equation [cf. Eq.~\eqref{eqn:excitonic_properties_1}].}
\label{table:data2}
\end{table}

\begin{figure}[t!]
    \centering
    \includegraphics[width=0.8\linewidth]{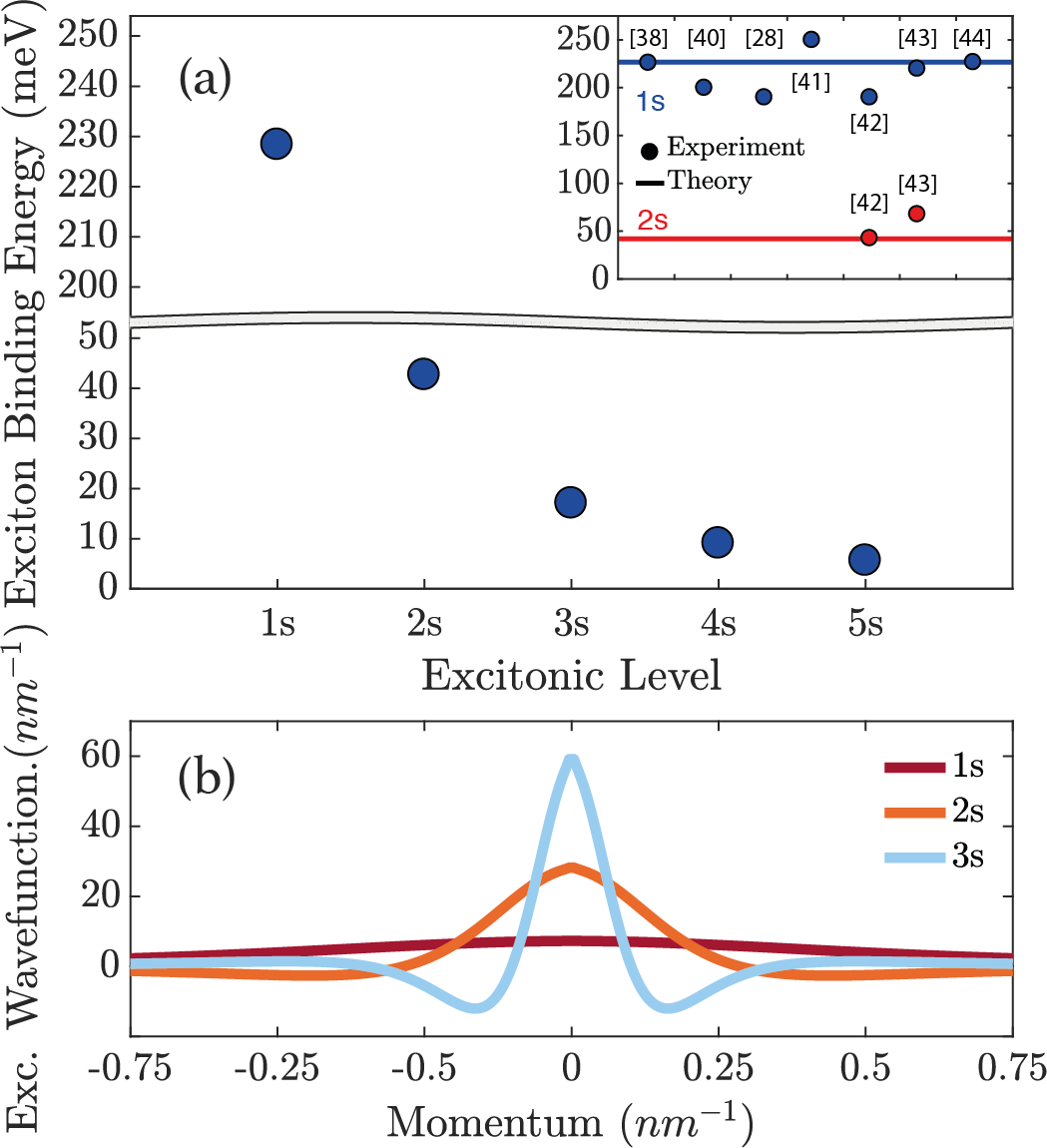}
    \caption{(a) Exciton binding energy of the energetically lowest s-type excitonic states in 2D perovskites. The inset shows a direct comparison to the measured values denoted by dots (and the corresponding references\,\cite{ishihara,gauthron,straus,cheng,zhai,urban2020revealing, delport2019exciton}). (b) Momentum-dependent wavefunctions for 1s, 2s, and 3s excitonic states, respectively.}
    \label{fig:exc properties}
\end{figure}

Evaluating the Wannier equation, we find $E_b^{\text{1s}}= 228$ meV for the energetically lowest 1s exciton state, while the binding energy decreases to 43 meV and 17 meV in the case of 2s and 3s states, respectively, cf. Fig.~\ref{fig:exc properties}(a) and Table \ref{table:data2}. The obtained exciton binding energies  are in good agreement with experimentally measured values, cf.  the inset of Fig.~\ref{fig:exc properties}(a). Note that interestingly the 2p state has a larger binding energy than the corresponding 2s state similarly to the case of TMD monolayers\,\cite{gunnar16}. The corresponding eigenfunctions of the three lowest excitonic states are illustrated in Fig. ~\ref{fig:exc properties}(b) and show the typical momentum-dependence well known from the hydrogen problem.

The obtained excitonic binding energies and wavefunctions will be used in the following to calculate the photoluminescence spectrum and the linewidths of excitonic transitions in the investigated 2D perovskite material.\\

\begin{figure*}[t!]
    \includegraphics[width=\textwidth]{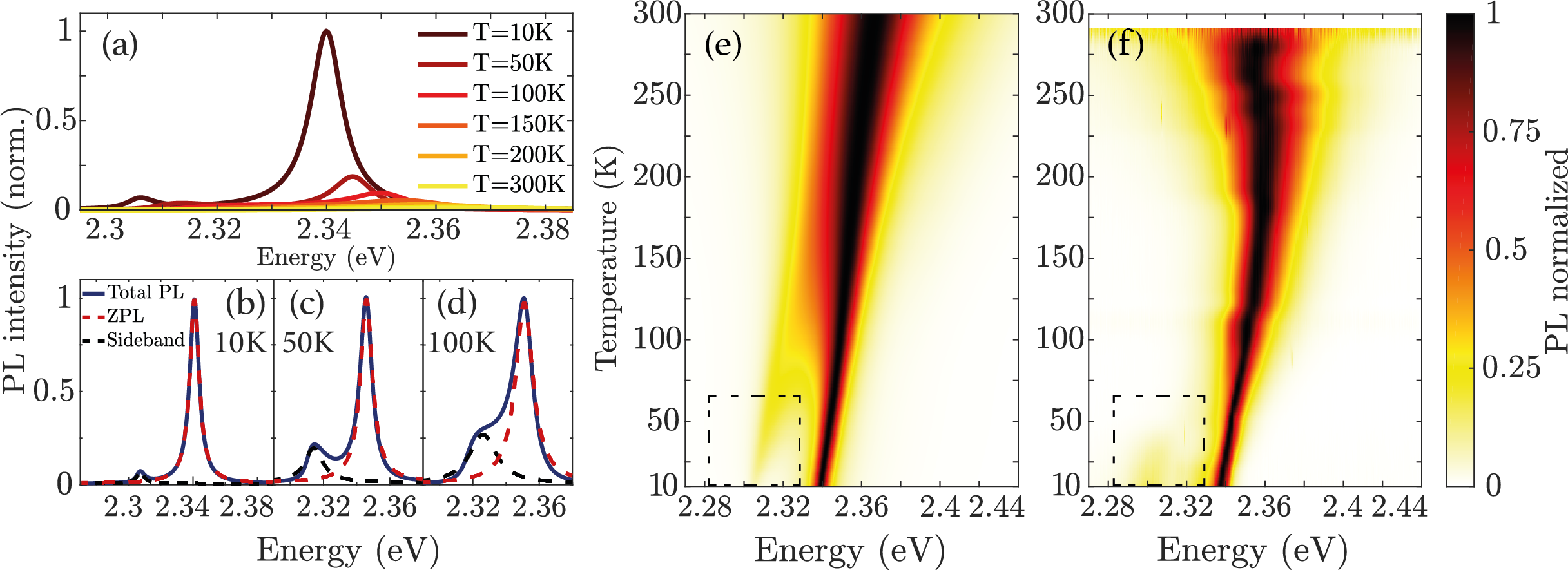}
    \caption{(a)  Photoluminescence spectra of 2D perovskites shown at fixed temperatures. (b)-(d) PL decomposed into contributions stemming from the zero-phonon line (ZPL, dashed red) and the phonon sideband (dashed black) at three specific temperatures. (e) Theoretical prediction and (f) experimental measurement of the temperature- and energy-resolved PL with the phonon sideband appearing at low temperatures (rectangular box).}
    \label{fig:PL properties}
\end{figure*}

The photoluminescence is calculated applying the density matrix formalism for an interacting system of electrons, phonons and
photons\,\cite{samuel20, carbonbuch, kadi2014}.
Using the Heisenberg equation of motion, we  calculate the temporal evolution of the photon number $n_\mathbf{k}=\langle c_\mathbf{k}^\dagger c_\mathbf{k} \rangle$, which determines the photoluminescence via $I_{PL}(E)=\dot{n}_\mathbf{k}$\,\cite{kira99, selig2018}. Here, radiative and non-radiative decay channels play the crucial role. The
radiative recombination of exciton states within the light cone is given by
$
\gamma_{0\sigma}^\nu=\frac{\hbar e_0^2}{2m_0^2 \varepsilon_0 n c_0}|M_\sigma|^2 \frac{|\phi_{\nu}(\mathbf{r}=0)|^2}{E_0^\nu}
$
with $E_0^{\nu}=E_g+E_b^{\nu}$ being the excitonic resonance energy and $M_{\sigma}$ being the optical matrix element for interband transitions projected into the polarization $\sigma$ of the emitted light \,\cite{selig2016,brem2019intrinsic}.
In the considered low-excitation regime, phonon-driven scattering channels determine the non-radiative dephasing\,\cite{selig2016,samuel20}
\begin{equation}
\label{eqn:PL1.2}
\Gamma^\nu_\mathbf{k}=\pi\sum_{j,\mathbf{q},\pm,\mu} |G^{\nu \mu j}_\mathbf{q}|^2\left(n^j_{\mathbf{q}}+\frac{1}{2}\pm\frac{1}{2}\right)\,\delta(\Delta E^{\mu \nu, j}_{\mathbf{k}, \mathbf{q}})
\end{equation}
with $\Delta E^{\mu \nu, j}_{\mathbf{k}, \mathbf{q}}=E^{\mu}_{\mathbf{k}+\mathbf{q}}-E^{\nu}_\mathbf{k} \pm \hbar \Omega^j_\mathbf{q}$. The excitonic dispersion $E^{\nu}_{\mathbf{k}}$ is described in an effective mass approximation with a total exciton mass $M = 0.44\ m_0$ \cite{ziegler2020fast}.
The equation includes absorption and emission ($\pm$) scattering channels involving acoustic and optical phonons phonons with the mode $j$. The emission term (+) includes spontaneous and stimulated processes. The latter is proportional to the phonon occupation  $n_\mathbf{q}^j$ that is given by the Bose-Einstein distribution within the considered bath approximation.

\begin{table}[b!]
\centering
\begin{tabular}{| c | c | c | c | c |}
 \hline
    $D_{AC}$ & $c_{AC}$ &$D_{OP}$ &  $\hbar \omega_{OP}$ &$\Gamma_0$ \\ [1ex]
 \hline
  1.9 eV & 2100 m/s & 158
eV/nm & 35 meV &  6.7 meV \\ [1ex]
 \hline
\end{tabular}
\caption{Values for the deformation potentials for optical and acoustic phonons are extracted from the performed measurements of temperature-dependent excitonic linewidths in PL spectra assuming one dominant optical and acoustic phonon branch.}
\label{table:data}
\end{table}

The appearing exciton-phonon matrix element
$
G^{\nu \mu j}_\mathbf{q}=\sum_{\mathbf{k},\alpha=e,h}g^{j,\alpha}_{\mathbf{q}} \phi_\mathbf{k}^{*\nu} \phi_{\mathbf{k}+\alpha\mathbf{q}}^{\mu}
$
 is given by the coupling elements for electrons and holes $g^{e/h}$ sandwiched by the exciton wavefunctions $\phi^\mu_\mathbf{k}$ of the involved initial and final exciton states. The electron-phonon coupling reads  $g^{j,\alpha}_{\mathbf{q}}=D_\mathbf{q}^{j,\alpha} \big(\frac{\hbar^2}{2\rho\hbar\omega^j_{\mathbf{q}}}\big)^{1/2}$ with a
 constant deformation potential for optical phonons, i.e.  $D_\mathbf{q}^{j,\alpha}=D_{OP}^{j,\alpha}$, and with $D_\mathbf{q}^{j,\alpha}=D_{AC}^{j,\alpha}|\mathbf{q}|$ for acoustic phonon modes.
First-principle studies are needed to obtain access to  the full phonon dispersion and electron-phonon coupling elements. While this is beyond the scope of this work, we extract the values for deformation potentials from the experimentally measured temperature-dependent linewidth of the 1s exciton assuming one dominant optical and acoustic phonon branch, cf. Table \ref{table:data}.  We find that the linear increase of the PL linewidth at low temperatures can be traced back to
the scattering with acoustic phonons characterized by the velocity $c_{AC}=2100$m/s. At high temperatures, there is a deviation from a linear increase and the  major contribution stems from scattering with optical phonons exhibiting an energy of 35 meV. Note that there is also a defect-induced linewidth $\Gamma_0 = 6.7$ meV at zero temperature.

Considering a many-particle Hamilton operator including exciton-photon and exciton-phonon interaction and evaluating the Heisenberg equation of motion for the  photon number $n_{\mathbf{k}}$, we find a coupled set of differential equations including phonon- and photon-assisted polarizations. Then, the cluster expansion scheme\,\cite{kira06} is used to factorize the many-particle expectation values disregarding contributions connected to multi-phonon processes \cite{samuel20}. The final analytic expression for the $\sigma$-polarized photon flux emitted in perpendicular direction with respect to the monolayer is given by\,\cite{samuel20}
\begin{eqnarray}
\label{eqn:pl}
&&I_{PL,\sigma}(E)=\frac{2}{\hbar} |M_{\sigma}|^2 \sum_\mu \frac{{|\phi^\mu(\mathbf{r}=0)|}^2} {(E_0^\mu-E)^2 + (\gamma^{\mu}_{0,\sigma}+\Gamma_0^\mu)^2} \times \\\nonumber
&&
\Big(\gamma^\mu_{0,\sigma} N_0^\mu +\sum_{\nu,\mathbf{q},j,\pm}|G_{\mathbf{q}}^{\nu \mu j}|^2N_\mathbf{q}^\nu \eta^{\pm}_{j\mathbf{q}}\frac{\Gamma^{\nu}_{\mathbf{q}}}{(E^{\nu}_{\mathbf{q}}\mp \Omega_{\mathbf{q}}^\alpha-E)^2+(\Gamma^{\nu}_{\mathbf{q}})^2}\Big)
\end{eqnarray}
where $\eta_{j\mathbf{q}}^{\pm}=1/2\pm1/2+n^j_\mathbf{q}$ denotes the relevant phonon
occupation factor for absorption/emission and $N_\mathbf{q}^\nu$ represents the exciton occupation corresponding to the Boltzmann distribution for the considered stationary PL. The first term  describes the direct photoluminescence that is often referred to as zero-phonon contribution. The second term shows the phonon-assisted indirect PL and results in the emergence of  phonon sidebands.
Evaluating Eq.~\eqref{eqn:pl}, we obtain a microscopic access to the temperature-dependent PL and find that the linewidth of the exciton transition clearly increases with temperature, while its intensity becomes smaller, cf. Fig.~\ref{fig:PL properties}(a). The zero-phonon linewidth ranges between $\sim5-10$ meV at low temperatures and $\sim50$ meV at room temperature.

In Figs.~\ref{fig:PL properties}(b)-(d), we decompose the direct PL stemming from radiative exciton recombination within the light cone ($\mathbf Q\approx 0$) from  the indirect contribution to the PL arising from phonon-assisted exciton recombination from states with a non-zero center-of-mass momentum. We find a clear separation of the two contributions up to 50 K. At higher temperatures, the zero-phonon line becomes so broad that the phonon sideband is only visible as a low-energy shoulder. It becomes evident in an asymmetrically broadened exciton transition toward lower energies. The energetic separation of $\sim35$ meV between the zero-phonon line and the phonon sideband is governed by the energy of the involved optical phonons. In contrast, the weak contribution of phonon absorption leads to a negligible indirect photoluminescence on the high energy side of the zero-phonon line.

Optical phonons also govern the broadening of the zero-phonon line for temperatures above 100K, while at lower temperature acoustic phonons dominate the PL linewidth, as will be further discussed later. The overall red-shift toward lower energies with decreasing temperature stems from a shift in the relative position of the conduction and valence bands due to polaronic effects\,\cite{christiansen2017} as well as due to the temperature-dependent elongation of the lattice. This phenomenon is known as Varshni shift and can be described with the empiric formula\,\cite{varshni}
$
E_g(T)=E_g(0)-\frac{\alpha_1 T^2}{\alpha_2+T},
$
where $E_g(0)$ is the band gap energy at T=0K and $\alpha_1,\alpha_2$ are material-specific constants. The red-shift is treated phenomenologically in this work and the material-specific constants are fitted to experimental data.

To be able to directly compare the theoretical prediction to experiment, we have also performed temperature-dependent PL
measurements of the same 2D perovskites. The
experimental PL spectra were collected from a monocrystalline
(PEA)$_2$PbI$_4$ thin sheet ($\sim 100$ nm) exfoliated from chemically
synthesized crystals. To provide better stability of the exfoliated
perovskite layer it was encapsulated by h-BN sheets (for more details
see Ref.\,\onlinecite{ziegler2020fast}). For temperature-dependent PL
measurements, the sample was mounted on the cold finger of a helium
flow cryostat with a quartz optical window. PL was excited with the
frequency-doubled output of a mode-locked Ti-sapphire laser and
tuned to 400 nanometers. The excitation laser beam was focused on
the sample using a $50\times$ microscope objective with a numerical
aperture of 0.55, giving a spot size of approximately one micrometer
diameter. The emitted PL was collected through the same objective
and directed to a spectrometer equipped with a liquid nitrogen
cooled charge-coupled device camera.

Figures~\ref{fig:PL properties}(e) and (f) present a comparison of theoretically predicted and experimentally measured temperature- and energy-dependent PL spectra.
 We find a good qualitative agreement in terms of the appearance of the phonon-sideband at the lower energy side of the zero-phonon line  (cf. the rectangular boxes). Both in experiment and theory, we find a pronounced phonon sideband appearing at 2.33 eV and staying clearly visible and distinguishable from the zero-phonon line for temperatures up to 50\,K. Above this temperature the phonon sideband visible in experiment seems to merge/vanish with the main peak in contrast to theoretical predictions where this occurs around 100\,K. We expect that this discrepancy can be related to h-BN encapsulation which might modify the motion of the inorganic layer and the related phonon mode  with respect to the bulk case\,\cite{straus}. Indeed in case of the bulk (PEA)$_2$PbI$_4$ crystals the phonon side-band can be resolved up to around 100\,K\,\cite{straus2019longer}. This observation may indicate that the electron-phonon interaction can be substantially modified in thin layers of 2D perovskites due to possible residual strain induced by the organic cation interacting with the encapsulating layer\,\cite{du2020stacking}. It is also worth to note that in encapsulated layers the LO phonon energy is slightly lower than in bulk samples\,\cite{straus, urban2020revealing, straus2019longer} which also points to the non-negligible impact of encapsulation.\\


The spectral broadening of excitonic resonances with temperature plays an important role for the visibility of phonon sidebands. Thus, we study the exciton-phonon scattering channels determining the temperature-dependent spectral linewidth of the three energetically lowest excitonic transitions. The linewidth is determined on a microscopic footing within the second-order Born-Markov approximation resulting in the momentum-dependent scattering rates given in Eq.~\eqref{eqn:PL1.2}.
The appearing sum over all phonon modes and momenta includes all scattering processes that fulfill the energy conservation, i.e. $
E^{\mu}_{\mathbf{k+q}}-E_{\mathbf k}^\nu \pm \hbar \Omega^j_\mathbf{q}=0
$, where $\Omega^j_\mathbf{q}$ is the energy of the phonon with the mode $j$ and momentum $\mathbf q$, while $E_{\mathbf k}^\nu$ and $E^{\mu}_{\mathbf{k+q}}$ are the energies of the initial and final excitonic states. We include absorption ($-$) and emission ($+$) processes with optical and acoustic phonons.
Since only the states within the light cone (i.e. $\mathbf k\approx 0$) contribute to the direct PL, the linewidth of the zero-phonon line is determined by the scattering rate $\Gamma^\nu_\mathbf{k\approx0}$, cf.  Fig.~\ref{fig:PEA n}(b).

 \begin{figure}[t!]
    \includegraphics[width=\linewidth]{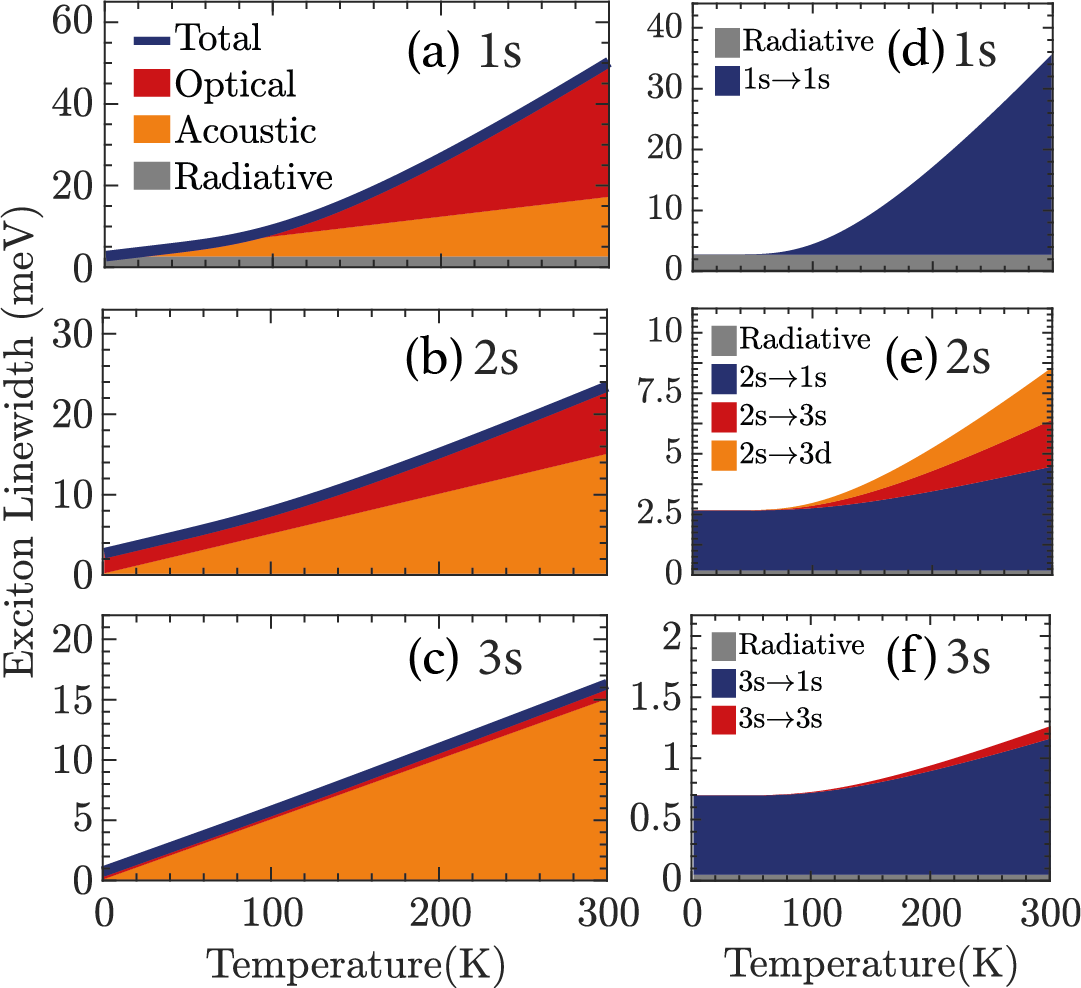}
    \caption{(a)-(c) Temperature-dependent exciton linewidth of the three energetically lowest excitonic transitions (1s, 2s and 3s) resolving the contribution of optical and acoustic phonons. (d)-(f) Microscopic scattering channels driven by optical phonons for the considered excitonic transitions. }
    \label{fig:lw properties}
\end{figure}

The resulting linewidths and their underlying microscopic scattering channels are displayed in Fig.~\ref{fig:lw properties}. We find spectral linewidths of approximately 50 meV, 25 meV and 15 meV for 1s, 2s and 3s transitions at room temperature, respectively. In the case of 1s, we show that the scattering with acoustic phonons dominates the linewidth up to temperatures of approximately 150 K, cf. the orange shaded region in Fig.~\ref{fig:lw properties}(a). We find a linear dependence on temperature reflecting the behaviour of the phonon occupations that are described by the Bose-Einstein distribution. Optical phonons start to be important for temperatures above 100 K (red region) and become the main channel governing the linewidth of 1s from approximately 150 K. Optical phonons are important only at higher temperatures, since their energy is 35 meV and thus the phonon occupation is negligibly small at low temperatures.
There is also a temperature-independent contribution stemming from radiative recombination (grey region, 2.66 meV) that also limits the lifetime of 1s excitons in the light cone. The oscillator strength considerably decreases for higher transitions due to more spatially extended wavefunctions. Hence, the radiative recombination has only a marginal contribution to their linewidths with $1.9 \cdot 10^{-1}\ \text{meV}$ for 2s and $4.9 \cdot 10^{-2}\ \text{meV}$ for 3s, cf. Figs.~\ref{fig:lw properties}(b)-(c). Note that we assume an ideal defect-free sample by disregarding the constant broadening $\Gamma_0$ that arises from scattering with defects and does not have any impact on the investigated temperature dependence.

\begin{figure}[t!]
    \centering
    \includegraphics[width=0.9\linewidth]{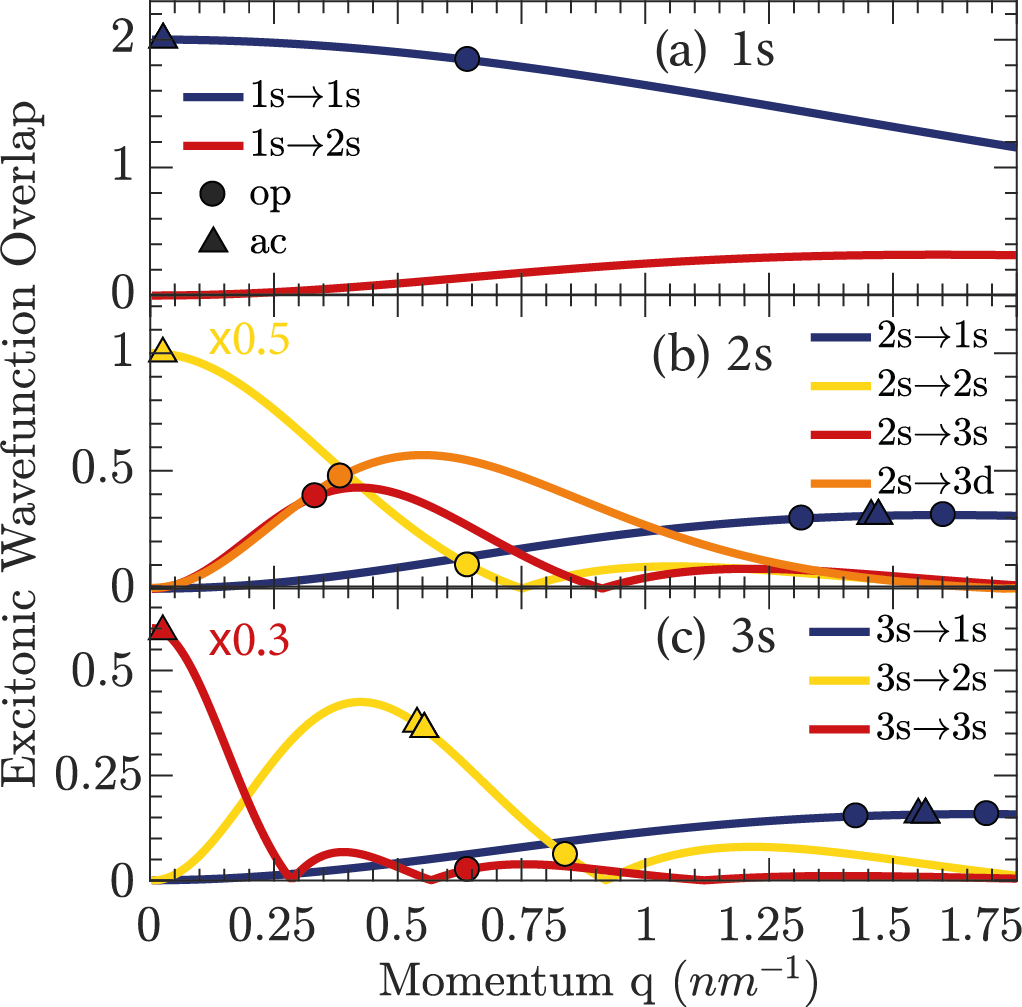}
    \caption{Wavefunction overlap of the excitonic states involved in phonon-driven transitions, which determine the spectral linewidth of (a) 1s, (b) 2s and (c) 3s states. The circular (triangular) markers indicate the momentum, at which the scattering can take place via optical (acoustic) phonons.  Here, the energy conservation is fulfilled. Note that the strongest wavefunction overlap involving the transitions within the same band have been multiplied by a factor of 0.5 or 0.3 to be able to better illustrate other scattering channels.}
    \label{fig:wf overlap}
\end{figure}

The quasi-linear dependence of the linewidth for higher states (2s and 3s) is caused by the fact that scattering with acoustic phonons is the main mechanism limiting the lifetime of 2s and 3s states at all temperatures.
To explain the overall low importance of optical phonons for higher excitonic states, we investigate the  wavefunction overlap $
\sum_{k,\alpha}\phi^{\nu*}_\mathbf{k} \phi^{\mu}_{\mathbf{k}+\alpha \mathbf{q}}
$ of the involved initial and final scattering states that enters in the exciton-phonon coupling elements.
In particular, it is important to consider the specific momenta where energy conservation is fulfilled, cf. circular and triangular markers  in Fig.~\ref{fig:wf overlap} denoting these momenta for optical and acoustic phonons respectively.
We find that the only possible scattering channel fulfilling the energy conservation for the 1s exciton is the intraband 1s$\rightarrow$1s scattering with optical and acoustic phonons, as shown in Fig.~\ref{fig:lw properties}(d).

For the higher 2s and 3s states, multiple scattering channels are possible, cf. Figs.~\ref{fig:wf overlap}(b)-(c).
Nevertheless, in all three cases it is the intraband scattering with acoustic phonons that shows the maximum wavefunction overlap due to the small momentum transfer of this transition. To be able to explain which specific scattering channel and phonon mode dominates the linewidth, it is not always sufficient to only consider the wavefunction overlap.  The exciton-phonon coupling element is determined by the wavefunction overlap weighted by the corresponding deformation potential that is constant for optical phonons and linear in $\mathbf{q}$ for acoustic phonons. Furthermore, the temperature- and energy-dependent occupation of the involved phonon modes also needs to be considered since it directly enters the scattering rate, cf. Eq.~\eqref{eqn:PL1.2}. The combination of a strong electron-phonon coupling, large wavefunction overlap, and a moderate phonon number results in a very efficient intraband scattering with acoustic phonons that dominates the linewidth. The efficiency of scattering with optical phonons decreases for higher states despite the enhanced number of possible scattering channels---as already seen in transition metal dichalcogenides \,\cite{brem2019intrinsic}. This can be traced back to lower excitonic binding energies and thus more spatially extended excitonic wavefunctions.

While acoustic phonons predominantly induce scattering within an excitonic band, optical phonons can also bridge larger energy distances between different bands. To better understand these processes, we further decompose the most important scattering channels with optical phonons in Figs.~\ref{fig:lw properties}(d)-(f). Since for optical phonons the coupling strength and phonon number do not depend on momentum, the efficiency of a particular scattering channel is only determined by the wavefunction overlap. We find that the most important contribution caused by optical phonons corresponds to emission processes to the energetically lower 1s state (blue areas in Figs\,\ref{fig:lw properties}(e)-(f)).
Note that spontaneous emission is much more likely than stimulated emission at low temperatures, where the phonon number is very small.   Moreover, the fact that spontaneous emission does not depend on the phonon number results in the constant linewidth at low temperatures.
For 2s excitons, the wavefunction overlap  for the optical phonon absorption processes 2s$\rightarrow$3s and 2s$\rightarrow$3d stands out (cf. Fig.~\ref{fig:wf overlap}(b)), which is directly reflected in their important contributions to the optical phonon portion of the linewidth in  Fig.~\ref{fig:lw properties}(e).

The 3s exciton state is close to the continuum allowing multiple scattering channels driven by emission and absorption of phonons.
The scattering with optical phonons at the 3s state is dominated by the 3s$\rightarrow$1s transition induced by spontaneous emission (cf. Fig.~\ref{fig:lw properties}(f)). Furthermore, due to the energetic proximity of other states, significant wavefunction overlap exists for interband acoustic phonon scattering processes (cf. Fig.~\ref{fig:wf overlap}(c)). However, the small phonon number at the large momenta involved in these scattering processes  drastically reduces the efficiency of these channels resulting in an even smaller linewidth at room temperature compared to the 1s exciton exhibiting much fewer scattering channels.\\

In summary, we have presented a theoretical study supported by experimental measurements shining light on excitonic properties of monolayered halide perovskites. Based on a quantum mechanical approach, we have determined the eigenfunctions and binding energies of the entire Rydberg-like series of excitonic states. Furthermore, we have calculated and measured temperature-dependent photoluminescence spectra with a particular focus on the emergence of low-energy phonon sidebands. Finally, we have provided microscopic insights into the exciton-phonon scattering channels governing the temperature-dependent spectral linewidth of the three energetically lowest excitonic transitions. Overall, our work contributes to a better understanding of optical properties of 2D perovskites and can guide future studies in this growing field of research.\\

\textbf{Acknowledgements:}
We thank Alexey Chernikov and Jonas D.  Ziegler (University of Regensburg) for providing the samples
and helping analyze the obtained results. Furthermore, we thank Samuel Brem (University of Chalmers) for discussions on modeling PL spectra. This project has received funding from the Swedish Research Council (VR, project number 2018-00734) and the European Union’s Horizon 2020 research and innovation programme under grant agreement no. 881603. The computations were enabled by resources provided by the Swedish National Infrastructure for Computing (SNIC) at C3SE partially funded by the Swedish Research Council through grant agreement no. 2016-07213. R.P.C. acknowledges funding from the Excellence Initiative Nano (Chalmers) under the Excellence PhD programme. P.P. appreciates support from National Science Centre Poland within the OPUS program (grant no. 2019/33/B/ST3/01915). This work was partially supported by OPEP project, which received funding from the ANR-10-LABX-0037-NEXT. M.D. appreciates support from the Polish National Agency for Academic Exchange within the Bekker programme (grant no. PPN/BEK/2019/1/00312/U/00001). K.W. and T.T. acknowledge support from the Elemental Strategy Initiative conducted by the MEXT, Japan, under Grant nos. JPMXP0112101001, JSPS KAKENHI, and JP20H00354, and the CREST (no. JPMJCR15F3), JST.\\

\textbf{Corresponding author:} causin@chalmers.se


\begin{thebibliography}{10}

\bibitem{jena19}
Ajay~Kumar Jena, Ashish Kulkarni, and Tsutomu Miyasaka.
\newblock Halide perovskite photovoltaics: Background, status, and future
  prospects.
\newblock {\em Chemical Reviews}, 119(5):3036--3103, 2019.

\bibitem{stoumpos13}
Constantinos~C. Stoumpos, Christos~D. Malliakas, and Mercouri~G. Kanatzidis.
\newblock Semiconducting tin and lead iodide perovskites with organic cations:
  Phase transitions, high mobilities, and near-infrared photoluminescent
  properties.
\newblock {\em Inorganic Chemistry}, 52(15):9019--9038, 2013.
\newblock PMID: 23834108.

\bibitem{niu15}
Guangda Niu, Xudong Guo, and Liduo Wang.
\newblock Review of recent progress in chemical stability of perovskite solar
  cells.
\newblock {\em J. Mater. Chem. A}, 3:8970--8980, 2015.

\bibitem{yin15}
Wan-Jian Yin, Ji-Hui Yang, Joongoo Kang, Yanfa Yan, and Su-Huai Wei.
\newblock Halide perovskite materials for solar cells: a theoretical review.
\newblock {\em J. Mater. Chem. A}, 3:8926--8942, 2015.

\bibitem{Tsai2016}
Hsinhan Tsai, Wanyi Nie, Jean-Christophe Blancon, Constantinos~C. Stoumpos,
  Reza Asadpour, Boris Harutyunyan, Amanda~J. Neukirch, Rafael Verduzco,
  Jared~J. Crochet, Sergei Tretiak, Laurent Pedesseau, Jacky Even, Muhammad~A.
  Alam, Gautam Gupta, Jun Lou, Pulickel~M. Ajayan, Michael~J. Bedzyk,
  Mercouri~G. Kanatzidis, and Aditya~D. Mohite.
\newblock High-efficiency two-dimensional ruddlesden--popper perovskite solar
  cells.
\newblock {\em Nature}, 536(7616):312--316, Aug 2016.

\bibitem{stoumpos14}
Feng Hao, Constantinos~C. Stoumpos, Robert P.~H. Chang, and Mercouri~G.
  Kanatzidis.
\newblock Anomalous band gap behavior in mixed sn and pb perovskites enables
  broadening of absorption spectrum in solar cells.
\newblock {\em Journal of the American Chemical Society}, 136(22):8094--8099,
  2014.
\newblock PMID: 24823301.

\bibitem{sampson}
Sampson Adjokatse, Hong-Hua Fang, and Maria~Antonietta Loi.
\newblock Broadly tunable metal halide perovskites for solid-state
  light-emission applications.
\newblock {\em Materials Today}, 20(8):413 -- 424, 2017.

\bibitem{nrel}
Best Research-Cell~Efficiencies NREL.

\bibitem{huang15}
Jinsong Huang, Yuchuan Shao, and Qingfeng Dong.
\newblock Organometal trihalide perovskite single crystals: A next wave of
  materials for 25\% efficiency photovoltaics and applications beyond?
\newblock {\em The Journal of Physical Chemistry Letters}, 6(16):3218--3227,
  2015.

\bibitem{Tan2014}
Zhi-Kuang Tan, Reza~Saberi Moghaddam, May~Ling Lai, Pablo Docampo, Ruben
  Higler, Felix Deschler, Michael Price, Aditya Sadhanala, Luis~M. Pazos, Dan
  Credgington, Fabian Hanusch, Thomas Bein, Henry~J. Snaith, and Richard~H.
  Friend.
\newblock Bright light-emitting diodes based on organometal halide perovskite.
\newblock {\em Nature Nanotechnology}, 9(9):687--692, Sep 2014.

\bibitem{Cho1222}
Himchan Cho, Su-Hun Jeong, Min-Ho Park, Young-Hoon Kim, Christoph Wolf,
  Chang-Lyoul Lee, Jin~Hyuck Heo, Aditya Sadhanala, NoSoung Myoung, Seunghyup
  Yoo, Sang~Hyuk Im, Richard~H. Friend, and Tae-Woo Lee.
\newblock Overcoming the electroluminescence efficiency limitations of
  perovskite light-emitting diodes.
\newblock {\em Science}, 350(6265):1222--1225, 2015.

\bibitem{Matsushima2019}
Toshinori Matsushima, Fatima Bencheikh, Takeshi Komino, Matthew~R. Leyden,
  Atula S.~D. Sandanayaka, Chuanjiang Qin, and Chihaya Adachi.
\newblock High performance from extraordinarily thick organic light-emitting
  diodes.
\newblock {\em Nature}, 572(7770):502--506, Aug 2019.

\bibitem{Gong2018}
Xiwen Gong, Oleksandr Voznyy, Ankit Jain, Wenjia Liu, Randy Sabatini, Zachary
  Piontkowski, Grant Walters, Golam Bappi, Sergiy Nokhrin, Oleksandr Bushuyev,
  Mingjian Yuan, Riccardo Comin, David McCamant, Shana~O. Kelley, and Edward~H.
  Sargent.
\newblock Electron--phonon interaction in efficient perovskite blue emitters.
\newblock {\em Nature Materials}, 17(6):550--556, Jun 2018.

\bibitem{stoumpos19}
Lingling Mao, Constantinos~C. Stoumpos, and Mercouri~G. Kanatzidis.
\newblock Two-dimensional hybrid halide perovskites: Principles and promises.
\newblock {\em Journal of the American Chemical Society}, 141(3):1171--1190,
  2019.
\newblock PMID: 30399319.

\bibitem{Mante2017}
Pierre-Adrien Mante, Constantinos~C. Stoumpos, Mercouri~G. Kanatzidis, and
  Arkady Yartsev.
\newblock Electron--acoustic phonon coupling in single crystal ch3nh3pbi3
  perovskites revealed by coherent acoustic phonons.
\newblock {\em Nature Communications}, 8(1):14398, Feb 2017.

\bibitem{lu17}
Xujie Lü, Wenge Yang, Quanxi Jia, and Hongwu Xu.
\newblock Pressure-induced dramatic changes in organic–inorganic halide
  perovskites.
\newblock {\em Chem. Sci.}, 8:6764--6776, 2017.

\bibitem{sun15}
Shijing Sun, Yanan Fang, Gregor Kieslich, Tim~J. White, and Anthony~K.
  Cheetham.
\newblock Mechanical properties of organic–inorganic halide perovskites{,}
  ch3nh3pbx3 (x = i{,} br and cl){,} by nanoindentation.
\newblock {\em J. Mater. Chem. A}, 3:18450--18455, 2015.

\bibitem{etgar18}
Lioz Etgar.
\newblock The merit of perovskite{'}s dimensionality; can this replace the 3d
  halide perovskite?
\newblock {\em Energy Environ. Sci.}, 11:234--242, 2018.

\bibitem{milot16}
Rebecca~L. Milot, Rebecca~J. Sutton, Giles~E. Eperon, Amir~Abbas Haghighirad,
  Josue Martinez~Hardigree, Laura Miranda, Henry~J. Snaith, Michael~B.
  Johnston, and Laura~M. Herz.
\newblock Charge-carrier dynamics in 2d hybrid metal–halide perovskites.
\newblock {\em Nano Letters}, 16(11):7001--7007, 2016.
\newblock PMID: 27689536.

\bibitem{yaffe15}
Omer Yaffe, Alexey Chernikov, Zachariah~M. Norman, Yu~Zhong, Ajanthkrishna
  Velauthapillai, Arend van~der Zande, Jonathan~S. Owen, and Tony~F. Heinz.
\newblock Excitons in ultrathin organic-inorganic perovskite crystals.
\newblock {\em Phys. Rev. B}, 92:045414, Jul 2015.

\bibitem{yaffe2015excitons}
Omer Yaffe, Alexey Chernikov, Zachariah~M Norman, Yu~Zhong, Ajanthkrishna
  Velauthapillai, Arend van~der Zande, Jonathan~S Owen, and Tony~F Heinz.
\newblock Excitons in ultrathin organic-inorganic perovskite crystals.
\newblock {\em Physical Review B}, 92(4):045414, 2015.

\bibitem{straus2019longer}
Daniel~B Straus, Natasha Iotov, Michael~R Gau, Qinghua Zhao, Patrick~J Carroll,
  and Cherie~R Kagan.
\newblock Longer cations increase energetic disorder in excitonic 2d hybrid
  perovskites.
\newblock {\em The journal of physical chemistry letters}, 10(6):1198--1205,
  2019.

\bibitem{ziegler2020fast}
Jonas~D Ziegler, Jonas Zipfel, Barbara Meisinger, Matan Menahem, Xiangzhou Zhu,
  Takashi Taniguchi, Kenji Watanabe, Omer Yaffe, David~A Egger, and Alexey
  Chernikov.
\newblock Fast and anomalous exciton diffusion in two-dimensional hybrid
  perovskites.
\newblock {\em Nano Letters}, 2020.

\bibitem{guo2016electron}
Zhi Guo, Xiaoxi Wu, Tong Zhu, Xiaoyang Zhu, and Libai Huang.
\newblock Electron--phonon scattering in atomically thin 2d perovskites.
\newblock {\em ACS nano}, 10(11):9992--9998, 2016.

\bibitem{ni2017real}
Limeng Ni, Uyen Huynh, Alexandre Cheminal, Tudor~H Thomas, Ravichandran
  Shivanna, Ture~F Hinrichsen, Shahab Ahmad, Aditya Sadhanala, and Akshay Rao.
\newblock Real-time observation of exciton--phonon coupling dynamics in
  self-assembled hybrid perovskite quantum wells.
\newblock {\em ACS nano}, 11(11):10834--10843, 2017.

\bibitem{gong2018electron}
Xiwen Gong, Oleksandr Voznyy, Ankit Jain, Wenjia Liu, Randy Sabatini, Zachary
  Piontkowski, Grant Walters, Golam Bappi, Sergiy Nokhrin, Oleksandr Bushuyev,
  et~al.
\newblock Electron--phonon interaction in efficient perovskite blue emitters.
\newblock {\em Nature materials}, 17(6):550--556, 2018.

\bibitem{gauthron2010optical}
K~Gauthron, JS~Lauret, L~Doyennette, G~Lanty, A~Al~Choueiry, SJ~Zhang,
  A~Brehier, L~Largeau, O~Mauguin, J~Bloch, et~al.
\newblock Optical spectroscopy of two-dimensional layered (c6h5c2h4-nh3) 2-pbi4
  perovskite.
\newblock {\em Optics express}, 18(6):5912--5919, 2010.

\bibitem{straus}
Daniel~B. Straus, Sebastian Hurtado~Parra, Natasha Iotov, Julian Gebhardt,
  Andrew~M. Rappe, Joseph~E. Subotnik, James~M. Kikkawa, and Cherie~R. Kagan.
\newblock Direct observation of electron–phonon coupling and slow vibrational
  relaxation in organic–inorganic hybrid perovskites.
\newblock {\em Journal of the American Chemical Society}, 138(42):13798--13801,
  2016.
\newblock PMID: 27706940.

\bibitem{palmieri2020mahan}
Tania Palmieri, Edoardo Baldini, Alexander Steinhoff, Ana Akrap, M{\'a}rton
  Koll{\'a}r, Endre Horv{\'a}th, L{\'a}szl{\'o} Forr{\'o}, Frank Jahnke, and
  Majed Chergui.
\newblock Mahan excitons in room-temperature methylammonium lead bromide
  perovskites.
\newblock {\em Nature Communications}, 11(1):1--8, 2020.

\bibitem{calabrese}
J.~Calabrese, N.~L. Jones, R.~L. Harlow, N.~Herron, D.~L. Thorn, and Y.~Wang.
\newblock Preparation and characterization of layered lead halide compounds.
\newblock {\em Journal of the American Chemical Society}, 113(6):2328--2330,
  1991.

\bibitem{smith14}
Ian~C. Smith, Eric~T. Hoke, Diego Solis-Ibarra, Michael~D. McGehee, and
  Hemamala~I. Karunadasa.
\newblock A layered hybrid perovskite solar-cell absorber with enhanced
  moisture stability.
\newblock {\em Angewandte Chemie International Edition}, 53(42):11232--11235,
  2014.

\bibitem{Koch2006}
S.~W. Koch, M.~Kira, G.~Khitrova, and H.~M. Gibbs.
\newblock Semiconductor excitons in new light.
\newblock {\em Nature Materials}, 5(7):523--531, Jul 2006.

\bibitem{brem2018}
Samuel Brem, Malte Selig, Gunnar Berghaeuser, and Ermin Malic.
\newblock Exciton relaxation cascade in two-dimensional transition metal
  dichalcogenides.
\newblock {\em Scientific reports}, 8(1):8238, 2018.

\bibitem{selig2018}
Malte Selig, Gunnar Bergh{\"a}user, Marten Richter, Rudolf Bratschitsch,
  Andreas Knorr, and Ermin Malic.
\newblock Dark and bright exciton formation, thermalization, and
  photoluminescence in monolayer transition metal dichalcogenides.
\newblock {\em 2D Materials}, 5(3):035017, 2018.

\bibitem{rytova1967}
N~S. Rytova.
\newblock The screened potential of a point charge in a thin film.
\newblock {\em Moscow University Physics Bulletin}, 3(3):18, 1967.

\bibitem{keldysh}
L.~V. {Keldysh}.
\newblock {Coulomb interaction in thin semiconductor and semimetal films}.
\newblock {\em Soviet Journal of Experimental and Theoretical Physics Letters},
  29:658, June 1979.

\bibitem{malic_keldysh}
Gunnar Bergh\"auser and Ermin Malic.
\newblock Analytical approach to excitonic properties of mos${}_{2}$.
\newblock {\em Phys. Rev. B}, 89:125309, Mar 2014.

\bibitem{ishihara}
X.~Hong, T.~Ishihara, and A.~V. Nurmikko.
\newblock Dielectric confinement effect on excitons in
  ${\mathrm{pbi}}_{4}$-based layered semiconductors.
\newblock {\em Phys. Rev. B}, 45:6961--6964, Mar 1992.

\bibitem{gunnar16}
Gunnar Berghäuser, Andreas Knorr, and Ermin Malic.
\newblock Optical fingerprint of dark 2p-states in transition metal
  dichalcogenides.
\newblock {\em 2D Materials}, 4(1):015029, dec 2016.

\bibitem{gauthron}
K.~Gauthron, J-S. Lauret, L.~Doyennette, G.~Lanty, A.~Al Choueiry, S.J. Zhang,
  A.~Brehier, L.~Largeau, O.~Mauguin, J.~Bloch, and E.~Deleporte.
\newblock Optical spectroscopy of two-dimensional layered (c6h5c2h4-nh3)2-pbi4
  perovskite.
\newblock {\em Opt. Express}, 18(6):5912--5919, Mar 2010.

\bibitem{cheng}
Bin Cheng, Ting-You Li, Partha Maity, Pai-Chun Wei, Dennis Nordlund, Kang-Ting
  Ho, Der-Hsien Lien, Chun-Ho Lin, Ru-Ze Liang, Xiaohe Miao, Idris~A. Ajia, Jun
  Yin, Dimosthenis Sokaras, Ali Javey, Iman~S. Roqan, Omar~F. Mohammed, and
  Jr-Hau He.
\newblock Extremely reduced dielectric confinement in two-dimensional hybrid
  perovskites with large polar organics.
\newblock {\em Communications Physics}, 1(1):80, 2018.

\bibitem{zhai}
Yaxin Zhai, Sangita Baniya, Chuang Zhang, Junwen Li, Paul Haney, Chuan-Xiang
  Sheng, Eitan Ehrenfreund, and Zeev~Valy Vardeny.
\newblock Giant rashba splitting in 2d organic-inorganic halide perovskites
  measured by transient spectroscopies.
\newblock {\em Science Advances}, 3(7), 2017.

\bibitem{urban2020revealing}
Joanna~M Urban, Gabriel Chehade, Mateusz Dyksik, Matan Menahem, Alessandro
  Surrente, Gaelle Trippe-Allard, Duncan~K Maude, Damien Garrot, Omer Yaffe,
  Emmanuelle Delporte, et~al.
\newblock Revealing excitonic phonon coupling in (pe) 2 (ma) n- 1pbni3n+ 12d
  layered perovskites.
\newblock {\em The Journal of Physical Chemistry Letters}, 2020.

\bibitem{delport2019exciton}
G{\'e}raud Delport, Gabriel Chehade, Ferdinand L{\'e}d{\'e}e, Hiba Diab, Cosme
  Milesi-Brault, Gaelle Trippe-Allard, Jacky Even, Jean-S{\'e}bastien Lauret,
  Emmanuelle Deleporte, and Damien Garrot.
\newblock Exciton--exciton annihilation in two-dimensional halide perovskites
  at room temperature.
\newblock {\em The Journal of Physical Chemistry Letters}, 10(17):5153--5159,
  2019.

\bibitem{samuel20}
Samuel Brem, August Ekman, Dominik Christiansen, Florian Katsch, Malte Selig,
  Cedric Robert, Xavier Marie, Bernhard Urbaszek, Andreas Knorr, and Ermin
  Malic.
\newblock Phonon-assisted photoluminescence from indirect excitons in
  monolayers of transition-metal dichalcogenides.
\newblock {\em Nano Letters}, 20(4):2849--2856, 2020.
\newblock PMID: 32084315.

\bibitem{carbonbuch}
Ermin Malic and Andreas Knorr.
\newblock {\em Graphene and Carbon Nanotubes: Ultrafast Optics and Relaxation
  Dynamics}.
\newblock John Wiley \& Sons, 2013.

\bibitem{kadi2014}
Faris Kadi, Torben Winzer, Ermin Malic, Andreas Knorr, F~G{\"o}ttfert,
  M~Mittendorff, S~Winnerl, and M~Helm.
\newblock Microscopic description of intraband absorption in graphene: the
  occurrence of transient negative differential transmission.
\newblock {\em Physical review letters}, 113(3):035502, 2014.

\bibitem{kira99}
M.~Kira, F.~Jahnke, W.~Hoyer, and S.W. Koch.
\newblock Quantum theory of spontaneous emission and coherent effects in
  semiconductor microstructures.
\newblock {\em Progress in Quantum Electronics}, 23(6):189 -- 279, 1999.

\bibitem{selig2016}
Malte Selig, Gunnar Bergh{\"a}user, Archana Raja, Philipp Nagler, Christian
  Sch{\"u}ller, Tony~F Heinz, Tobias Korn, Alexey Chernikov, Ermin Malic, and
  Andreas Knorr.
\newblock Excitonic linewidth and coherence lifetime in monolayer transition
  metal dichalcogenides.
\newblock {\em Nature communications}, 7:13279, 2016.

\bibitem{brem2019intrinsic}
Samuel Brem, Jonas Zipfel, Malte Selig, Archana Raja, Lutz Waldecker, Jonas~D
  Ziegler, Takashi Taniguchi, Kenji Watanabe, Alexey Chernikov, and Ermin
  Malic.
\newblock Intrinsic lifetime of higher excitonic states in tungsten diselenide
  monolayers.
\newblock {\em Nanoscale}, 11(25):12381--12387, 2019.

\bibitem{kira06}
M.~Kira and S.W. Koch.
\newblock Many-body correlations and excitonic effects in semiconductor
  spectroscopy.
\newblock {\em Progress in Quantum Electronics}, 30(5):155 -- 296, 2006.

\bibitem{christiansen2017}
Dominik Christiansen, Malte Selig, Gunnar Bergh{\"a}user, Robert Schmidt, Iris
  Niehues, Robert Schneider, Ashish Arora, Steffen~Michaelis de~Vasconcellos,
  Rudolf Bratschitsch, Ermin Malic, et~al.
\newblock Phonon sidebands in monolayer transition metal dichalcogenides.
\newblock {\em Physical review letters}, 119(18):187402, 2017.

\bibitem{varshni}
Y.P. Varshni.
\newblock Temperature dependence of the energy gap in semiconductors.
\newblock {\em Physica}, 34(1):149 -- 154, 1967.

\bibitem{du2020stacking}
Qin Du, Cheng Zhu, Zixi Yin, Guangren Na, Chuantong Cheng, Ying Han, Na~Liu,
  Xiuxiu Niu, Huanping Zhou, Hongda Chen, et~al.
\newblock Stacking effects on electron--phonon coupling in layered hybrid
  perovskites via microstrain manipulation.
\newblock {\em ACS nano}, 14(5):5806--5817, 2020.

\end{thebibliography}

\end{document}